\documentclass[aps,prl,twocolumn]{revtex4}

\usepackage{latexsym,amsmath,graphicx,epsfig}
\usepackage{bm} 

\begin{document}

\title{Kerr Black Holes are Not Unique to General Relativity}

\author{Dimitrios Psaltis$^{1,2}$, Delphine Perrodin$^1$,
Keith R. Dienes$^1$, and Irina Mocioiu$^{3}$}
\affiliation{$^1$Department of Physics, 
University of Arizona, Tucson, AZ 85721 \\ 
$^2$Department of Astronomy, University of Arizona, Tucson, AZ 85721 \\
$^3$Department of Physics, Pennsylvania State University, University
Park, PA 16802}

\begin{abstract}
Considerable attention has recently focused on gravity theories
obtained by extending general relativity with additional scalar,
vector, or tensor degrees of freedom.  In this paper, we show that the
black-hole solutions of these theories are essentially
indistinguishable from those of general relativity. Thus, we conclude
that a potential observational verification of the Kerr metric around
an astrophysical black hole cannot, in and of itself, be used to
distinguish between these theories. On the other hand, it remains true
that detection of deviations from the Kerr metric will signify the
need for a major change in our understanding of gravitational physics.
\end{abstract}

\maketitle

Black holes are among the most extreme astrophysical objects predicted
by general relativity. They are vacuum solutions of the Einstein field
equations realized astrophysically at the end stages of the collapse
of massive stars. According to a variety of no-hair theorems, a
general relativistic black hole is characterized only by three
parameters identified with its gravitational mass, spin, and
charge. Any additional ``hair'' on the black hole, associated with the
properties of the progenitor star or the collapse itself, are radiated
away in the form of gravitational waves over a finite amount of time.

Black holes might look different if general relativity is only an
effective theory of gravity, valid at the curvature scales probed by
current terrestrial and astrophysical experiments. If the more
fundamental gravity theory has additional degrees of freedom, they
might appear as additional ``hair'' to the black hole. This would be
important for a number of reasons. First, additional degrees of
freedom appear naturally in all attempts to quantize gravity, either
in a perturbative approach~\cite{QGR} or within the context of string
theory~\cite{string}. Detecting observational signatures of these additional
degrees of freedom in black-hole spacetimes would serve as a confirmation
of quantum gravity effects. Second, black-hole solutions not described by
the Kerr-Newman metric may follow a set of thermodynamic relations
different than those calculated by Bekenstein~\cite{bek} and
Hawking~\cite{hawk} with important implications for string
theory~\cite{malda}. Finally, the external spacetimes of astrophysical
black hole will soon be mapped with gravitational-wave~\cite{bh_gw}
and high-energy observations~\cite{bh_xray} and the means for
searching for black holes with additional degrees of freedom will
become readily available.

Introducing additional degrees of freedom to the Einstein-Hilbert
action of the gravitational field does not necessarily alter the
resulting field equations and hence the black-hole solutions. For
example, the addition of a Gauss-Bonnet term to the action leaves the
field equation completely unchanged~\cite{QGR}. Moreover, a large
class of gravity theories in the Palatini formalism for which the
action is a general function $f(R)$ of the Ricci scalar curvature $R$,
lead to field equations that are indistinguishable from the general
relativistic ones~\cite{fRPalatini}. In all these situations, no
astrophysical observation of a classical phenomenon, such as test
particle orbits or gravitational lensing, can distinguish between these
theories. Nevertheless, this leaves a large number of Lagrangian gravity
theories that incorporate general relativity as a limiting case but 
are described by more general field equations.

The most widely studied such extension of general relativity is the
Brans-Dicke gravity, which incorporates a dynamical scalar field in
addition to the metric tensor. Black hole solutions in this theory
were studied by Thorne \& Dykla~\cite{TD71}. Following a conjecture by
Penrose, these authors showed that the Kerr solution of general
relativity is also an exact solution of the field equations in
Brans-Dicke gravity and offered a number of arguments to support the
claim that the collapse of a star in this gravity theory will produce
uniquely a Kerr black hole. Additional
analytic~\cite{hawk,BDBH_bek,BDBH} and numerical~\cite{numerical}
arguments were offered by other authors providing further evidence for
the uniqueness of the Kerr solution in Brans-Dicke gravity.  

In this Letter, we show that black-hole solutions of the general
relativistic field equations are indistinguishable from solutions of a
wide variety of gravity theories that arise by adding dynamical vector
and tensor degrees of freedom to the Einstein-Hilbert action. Although
we do not prove that the general relativistic vacuum solution is the
{\em unique\/} solution of the extended Lagrangian theories, we use
our results to argue that an observational verification of the Kerr
solution for an astrophysical object cannot be used in distinguishing
between general relativity and other Lagrangian theories such as those
considered here. Note that we are only considering four-dimensional
theories that obey the equivalence principle, and hence we are not
studying theories with prior geometry~\cite{Rosen}, that are Lorentz
violating~\cite{Lorviol}, or braneworld gravity theories~\cite{Brane}.
Although several of these extensions lead to predictions of an unstable
quantum vacuum and of ghosts, we are focusing here on their classical
black-hole solutions.

In general relativity, the external spacetimes of black holes that are
astrophysically relevant, i.e., with zero charge, are completely
specified by the relation
\begin{equation}
R_{\mu\nu}=\frac{R}{4} g_{\mu\nu}\;,
\label{eq:GRsol}
\end{equation}
with $R_{,\mu}=0$. Here $R_{\mu\nu}$ is the Ricci tensor and $R$
is the Ricci scalar curvature. When the cosmological constant
$\Lambda$ is considered to be non-zero, then $R=4\Lambda$.

It is our aim to show that the external spacetimes of general
relativistic black holes, which satisfy equation~(\ref{eq:GRsol}), are
practically indistinguishable from solutions in a number of gravity
theories that arise by adding vector or tensor degrees of freedom
to the Einstein-Hilbert action
\begin{equation}
S= \frac{1}{16\pi G} \int d^4x\sqrt{-g}\left(-2\Lambda+ R\right)\;.
\label{eq:EH}
\end{equation}

It is important to make here a distinction between the external
spacetimes of black holes and those of stellar objects, which also
satisfy equation~(\ref{eq:GRsol}) in general relativity. The field
equation of a gravity theory is a high-order partial differential
equation and its solutions depend on the boundary conditions
imposed. In particular, when solving for the external spacetime of a
stellar object, a number of regularity conditions need to be satisfied
at the stellar surface, inside which the field equations are altered
by the presence of matter. As a result, proving that the external
spacetime of a general relativistic star satisfies the vacuum field
equation of a different gravity theory is not a guarantee that it will
be a valid solution for that theory, as well. It also needs to meet
the altered regularity conditions at the stellar surface. This issue
was recently explored for $1/R$ gravity in the
metric~\cite{boundary_metric} and in the Palatini
formalism~\cite{boundary_Pal} with important implications for the
viability of this theory. This concern, however, is not relevant for
black-hole solutions, in which there is no matter anywhere outside the
horizon and hence no regularity conditions need to be met. Indeed the
vacuum field equation is valid throughout the entire spacetime
accessible to a distant observer and only the boundary conditions at
radial infinity need to be checked.

\noindent{\em $f(R)$ Gravity in the Metric Formalism.---\/}
A self-consistent theory of gravity can be constructed for any
Lagrangian action that obeys a small number (four) of simple
requirements~\cite{MTW}. Of all the possibilities, the field equations
that are derived from the Einstein-Hilbert action~(\ref{eq:EH}) are
the only ones that are also linear in the Riemann tensor and result in
field equations that are of second order. However, any other action
$f(R)$ that depends only on the Ricci curvature scalar will also
satisfy the above four requirements~\cite{Wald}, while being free of
the Ostrogradski instability~\cite{Wood}.

The field equation that results from extremizing an action that is a
general function of the Ricci scalar, $f(R)$, is
\begin{eqnarray}
& &\left(-R_{;k}R_{;l}+g_{kl}R_{;m}R^{;m}\right)f^{\prime\prime\prime}(R) 
  \nonumber\\
& &\qquad\qquad
+\left(-R_{;kl}+g_{kl}\Box R\right)f^{\prime\prime}(R)\nonumber\\
& & \qquad\qquad\qquad
  +R_{kl}f^\prime(R)-\frac{1}{2}g_{kl}f(R)=0\;,
\label{eq:fR}
\end{eqnarray}
where primes denote differentiation with respect to $R$ and we have used
the sign convention of Ref.~\cite{MTW}.

A general relativistic black-hole solution, i.e., one that satisfies
equation~(\ref{eq:GRsol}) with $R_{,\mu}=0$, will also be a solution
of the field equation~(\ref{eq:fR}) if
\begin{equation}
\frac{1}{2}Rf^\prime(R)-f(R)=0\;.
\label{eq:trace}
\end{equation}
We will now consider non-pathological functional forms of $f(R)$ that
can be expanded in a Taylor series of the form
\begin{equation}
f(R)=a_0+R+a_2R^2+a_3R^3+...a_nR^n+...\;,
\label{eq:taylor}
\end{equation}
where we have normalized all coefficients with respect to the
coefficient of the linear term. The Einstein-Hilbert action is the
specific case of equation~(\ref{eq:taylor}) for $a_0=-2\Lambda$, and
$a_{n\ge 2}=0$. We can then write the condition~(\ref{eq:trace}) for the
existence of a constant curvature solution as
\begin{equation}
-a_0-\frac{1}{2}R +\frac{1}{2}a_3 R^3 + ...
+ \frac{n-2}{2}a_n R^n + ...=0\;.
\label{eq:fRcond}
\end{equation}

There are three cases to consider: {\em (i)\/} If $a_0=0$, then the
Kerr solution, which corresponds to $R=0$, will always be a solution
of the field equations of a general $f(R)$ theory.  Thus, in the
absence of a cosmological constant, we conclude that the Kerr solution
of general relativity remains an exact solution to all $f(R)$ theories
as long as $f(R)$ has a Taylor expansion of the form in
Eq.~(\ref{eq:taylor}).

{\it (ii)} Moreover, independent of the value of $a_0$, all of the
constant-curvature solutions of General Relativity in vacuum --
including the Kerr solution -- remain exact solutions of the $f(R)$
theory, if the Taylor series for $f(R)$ terminates after the quadratic
term (i.e., if $a_{n\geq 3}=0$). Indeed, this statement remains true
independently of the value of $a_0$, and thus holds for both vanishing
and non-vanishing cosmological constants.

{\it (iii)} Finally, if $a_0\ne 0$ and the Taylor expansion extends
beyond the quadratic term, then Kerr-like black-hole solution will
always be possible.  The only change is that the value of its constant
curvature will be shifted relative to the value predicted in General
Relativity.  Since terrestrial and solar-system tests require any
extra non-linear terms in the gravity action to be perturbative, this
shift in the curvature will also be correspondingly small.  However,
even in this case, it is straightforward to show that the corrections
to the curvature are actually suppressed by additional powers of the
cosmological constant relative to what might naively have been
expected on the basis of dimensional analysis.  For example, given the
expansion for $f(R)$ in Eq.(\ref{eq:taylor}), we would have expected
the curvature term to have a leading correction term which scales as
$R = -2 a_0[1 + {\cal O}(a_0 a_2) + ...]$.
However, explicitly solving Eq.~(\ref{eq:fRcond}), we find that the
true leading correction is actually given by
\begin{equation}
R=-2 a_0 \left(1+4a_0^2 a_3+...\right)\;.
\label{eq:concurv}
\end{equation}
Thus the deviations of the vacuum curvature solutions of $f(R)$
gravity from those of General Relativity are particularly suppressed.

\noindent{\em $f(R)$ Gravity in the Palatini Formalism.---\/}
In deriving the field equation~(\ref{eq:fR}), we extremized the action
of the gravitational field with respect only to variations in the
metric.  In the so-called Palatini formalism, field equations of lower
order can be derived from the same action of the gravitational field,
by extremizing it over both the metric and the
connection~\cite{Sotiriou}. A large class of $f(R)$ theories in the
Palatini formalism are known to result in the same field equations as
general relativity~\cite{fRPalatini}.

Applying this procedure for a gravitational action that is a general
function $f(R)$ of the Ricci scalar curvature, we obtain the
well-known set of equations~\cite{Sotiriou}
\begin{eqnarray}
R_{\rm kl} f^\prime (R)-\frac{1}{2}g_{kl} f(R)&=&0
\label{eq:Pal}\\
\nabla_\sigma \left[ \sqrt{-g} f^\prime(R) g^{\mu\nu}\right]=0
\label{eq:Conn}\;.
\end{eqnarray}

In order to look for constant curvature solutions in vacuum for this
theory, we first take the trace of equation~(\ref{eq:Pal}). The result
is simply the algebraic equation~(\ref{eq:trace}), which we can solve
for the value of the constant curvature~(\ref{eq:concurv}) as before.
For a solution with constant curvature, the factor $f^\prime(R)$ in
equation~(\ref{eq:Conn}) is a constant, and the solutions to this
equation are simply the Christoffel symbols of general relativity.  As
a result, any general relativistic solution of constant curvature,
such as the black-hole solutions with cosmological constant, will also
be solutions (with the same or slightly different value of the
cosmological constant) to the field equations of an $f(R)$ gravity in
the Palatini formalism.

\noindent{\em General Quadratic Gravity.---\/}
We shall now consider a gravitational action that incorporates all
combinations of the Ricci curvature, Ricci tensor, and Riemann tensor,
up to second order, i.e.,
\begin{eqnarray}
S&=& \frac{1}{16\pi G} \int d^4x\sqrt{-g}\left(-2\Lambda+ R + \alpha R^2 + 
     \beta R_{\sigma\tau} R^{\sigma\tau} \right.\nonumber\\
 & &\qquad\qquad\qquad \left. + 
 \gamma R_{\alpha\beta\gamma\delta} R^{\alpha\beta\gamma\delta} \right)\;.
\label{eq:ac2general}
\end{eqnarray}
with $\alpha$, $\beta$, and $\gamma$ the parameters of the theory.
Such terms appear naturally as radiative corrections to the
Einstein-Hilbert action in perturbative approaches to quantum
gravity~\cite{QGR} or in string theory~\cite{string}. Note, however,
that in general such theories are not free of the Ostrogradski
instability~\cite{Wood}.

Because of the Gauss-Bonnet identity, the predictions of the theory
described by the action~(\ref{eq:ac2general}) in calculating classical
properties of astrophysical black holes are identical to those of the
action~\cite{barrow}
\begin{equation}
S=\frac{1}{16\pi G} \int{\sqrt{-g}\left(-2\Lambda+ R + \alpha' R^2 + 
    \beta' R_{\sigma\tau} R^{\sigma\tau} \right)},
\end{equation}
where $\alpha'=\alpha - \gamma$ and $\beta'=\beta + 4 \gamma$.

The field equation for this action in the metric formalism is
\begin{equation}
R_{\mu\nu}-\frac{1}{2} R g_{\mu\nu}+\alpha'
   K_{\mu\nu}+\beta' L_{\mu \nu} +\Lambda g_{\mu\nu} = 0
\label{eq:quadratic}
\end{equation}
where
\begin{eqnarray}
K_{\mu\nu} &\equiv& - 2 R_{;\mu\nu}+2 g_{\mu\nu} \Box R 
-\frac{1}{2} R^2 g_{\mu\nu} + 2 R R_{\mu \nu},\\
L_{\mu\nu} &\equiv& - 2  R_{\mu \, \, \, \, ; \sigma \nu}^{\, \, \, 
\sigma} 
+ \Box R_{\mu\nu} + \frac{1}{2} g_{\mu\nu}  \Box R - \nonumber\\
& &\qquad\qquad
\frac{1}{2} \,g_{\mu\nu} \, R_{\sigma\tau}  R^{\sigma\tau}+2 
 R_{\mu}^{\ \alpha}  R_{\alpha\nu}\;.
\end{eqnarray}

It is trivial to show that, for any black-hole solution satisfying
equation~(\ref{eq:GRsol}), $K_{\mu\nu}=L_{\mu\nu}=0$ and the field equation of
quadratic gravity reduces to that of general relativity. As a result,
the Kerr solution is also a solution of the general quadratic theory
considered here.

\noindent{\em Vector-Tensor Gravity.---\/}
We finally consider a gravitational theory that incorporates a
dynamical vector field in addition to the metric tensor. A priori,
such an addition to the Einstein-Hilbert action appears to have the
highest probability of requiring black-hole solutions that are not
described by the Kerr metric. This is because the vector field has the
same spin as photons, the geodesics of which are used to define the
event horizon of a black hole. We restrict our attention to Lagrangian
theories that are linear and at most of second-order in the vector
field. The most general action for such a theory is~\cite{Willbook}
\begin{eqnarray}
S&=&\frac{1}{16\pi G}\int d^4x \sqrt{-g}\left(-2\Lambda+ R +\omega R K_\mu K^\mu 
   \right.\nonumber\\
& & \left.   + \eta K^\mu K^\nu R_{\mu\nu}
-\epsilon F_{\mu\nu}F^{\mu\nu}+\tau K_{\nu;\mu}K^{\mu;\nu}\right)\;,
\label{eq:vectens}
\end{eqnarray}
with
\begin{equation}
F_{\mu\nu}=K_{\nu;\mu}-K_{\mu;\nu}\;.
\end{equation}
The vector field $K_\mu$ at large distances from an object is meant to
asymptote smoothly to a background value determined by a cosmological
solution. Note that the values of the model parameters $\omega$,
$\eta$, $\epsilon$, and $\tau$ are not independent~\cite{Willbook}.

As in the case of previous investigations of scalar-tensor
gravity~\cite{TD71}, we will be seeking vacuum solutions that are
characterized by constant curvature, as well as by a constant vector
$K_\mu$. In this case, the field equations that are derived from the
action~(\ref{eq:vectens}) are~\cite{Willbook}
\begin{eqnarray}
R_{\mu\nu}-\frac{1}{2}Rg_{\mu\nu}+\omega\Theta^{(\omega)}_{\mu\nu}
+\eta \Theta^{(\eta)}_{\mu\nu}+\Lambda g_{\mu\nu}&=&0
   \label{eq:vectens1}\\
\omega K_\mu R + \eta K^\alpha R_{\mu\alpha}&=&0\label{eq:vectens2}\;,
\end{eqnarray}
where $K^2\equiv K_\mu K^\mu$,
\begin{eqnarray}
\Theta^{(\omega)}_{\mu\nu}&=&K_\mu K_\nu R + K^2 R_{\mu\nu}
-\frac{1}{2}g_{\mu\nu}K^2 R\;,\\
\Theta^{(\eta)}_{\mu\nu}&=&2K^\alpha K_\mu R_{\nu\alpha}-
2K^\alpha K_\nu R_{\mu\alpha}\nonumber\\
& & \qquad\qquad-\frac{1}{2}g_{\mu\nu}K^\alpha K^\beta
R_{\alpha\beta}\;.
\end{eqnarray}

We now multiply equation~(\ref{eq:vectens2}) by $K_\nu$, combine it
with equation~(\ref{eq:vectens1}), and look for the constant curvature
solution~(\ref{eq:GRsol}) to obtain
\begin{eqnarray}
&&\left[\Lambda-\frac{R}{4}\left(1+\omega K^2\right)\right]g_{\mu\nu}
    \nonumber\\
&&\qquad\qquad-\eta\frac{R}{4}\left(K_\mu K_\nu+
    \frac{1}{2}K^2g_{\mu\nu}\right)=0\;.
\label{eq:intermed}
\end{eqnarray}

Contracting equation~(\ref{eq:intermed}) with $g^{\mu\nu}$, we obtain
for the constant curvature
\begin{equation}
R=\frac{16\Lambda}{4+(4\omega+3\eta) K^2}
  \simeq 4\Lambda\left[1-\left(\omega+\frac{3\eta}{4}\right)K^2\right]\;.
\end{equation}
As in the previous cases, a black-hole solution that differs only in
the value of the constant curvature from the general relativistic one
is possible for the vector-tensor gravity theory that we have
considered.

\noindent{\em Discussion.---\/}
Our results have important implications for current attempts to test
general relativity in the strong-field regime using astrophysical
black holes. On the one hand, we appear to be lacking a parametric
theoretical framework with which to interpret observational data and
quantify possible deviations from the general relativistic predictions
for astrophysical black holes. On the other hand, the detection of
deviations from the Kerr metric in the spacetime of an astrophysical
black hole will be a very strong indication for the need of a major
change in our understanding of gravitation.

\bigskip

DP is supported in part by the Astrophysics Theory Program of NASA
under Grant NAG~513374.  KRD is supported in part by the U.S. National
Science Foundation under Grant PHY/0301998, by the U.S. Department of
Energy under Grant~DE-FG02-04ER-41298, and by a Research Innovation
Award from the Research Corporation. IM is supported by NSF grant
PHY-0555368. We are happy to thank Z.\ Chacko, C.-K.\ Chan, K.\
Sigurdson, and U.\ van Kolck for discussions.

\end{document}